\documentclass[reprint,superscriptaddress,amsmath,amssymb,aps,prd]{revtex4-1}
\usepackage[pdftex]{graphicx}
\usepackage{comment}
\usepackage{url}
\usepackage[labelformat=empty,labelsep=none]{caption}
\usepackage{bm}
\usepackage{comment}

\begin{document}

\title{Photoneutron Detection in Lightning by Gadolinium Orthosilicate Scintillators}

\author{Y.~Wada}
\affiliation{Department of Physics, Graduate School of Science, The University of Tokyo, 7-3-1 Hongo, Bunkyo-ku, Tokyo 113-0033, Japan}
\affiliation{High Energy Astrophysics Laboratory, Nishina Center for Accelerator-Based Science, RIKEN, 2-1 Hirosawa, Wako, Saitama 351-0198, Japan}
\author{K.~Nakazawa}
\affiliation{Kobayashi-Maskawa Institute for the Origin of Particles and the Universe, Nagoya University, Furo-cho, Chikusa-ku, Nagoya, Aichi 464-8601, Japan}
\author{T.~Enoto}
\affiliation{Extreme Natural Phenomena RIKEN Hakubi Research Team, Cluster for Pioneering Research, RIKEN, 2-1 Hirosawa, Wako, Saitama 351-0198, Japan}
\author{Y.~Furuta}
\affiliation{Collaborative Laboratories for Advanced Decommissioning Science, Japan Atomic Energy Agency, 2-4 Shirane Shirakata, Tokai-mura, Naka-gun, Ibaraki 319-1195, Japan}
\author{T.~Yuasa}
\affiliation{Block 4B, Boon Tiong Road, Singapore 165004, Singapore}
\author{K.~Makishima}
\affiliation{Department of Physics, Graduate School of Science, The University of Tokyo, 7-3-1 Hongo, Bunkyo-ku, Tokyo 113-0033, Japan}
\affiliation{High Energy Astrophysics Laboratory, Nishina Center for Accelerator-Based Science, RIKEN, 2-1 Hirosawa, Wako, Saitama 351-0198, Japan}
\affiliation{Kavli Institute for the Physics and Mathematics of the Universe, The University of Tokyo, 5-1-5 Kashiwa-no-ha, Kashiwa, Chiba 277-8683, Japan}
\author{H.~Tsuchiya}
\affiliation{Nuclear Science and Engineering Center, Japan Atomic Energy Agency, 2-4 Shirane Shirakata, Tokai-mura, Naka-gun, Ibaraki 319-1195, Japan}

\begin{abstract}

	During a winter thunderstorm on November 24, 2017, a downward terrestrial gamma-ray flash 
	took place and triggered photonuclear reactions with atmospheric nitrogen and oxygen nuclei,
	coincident with a lightning discharge at the Kashiwazaki-Kariwa nuclear power station in Japan.
	We directly detected 	neutrons produced by the photonuclear reactions with gadolinium orthosilicate scintillation crystals installed at sea level.
	Two gadolinium isotopes included in the scintillation crystals, $^{155}$Gd and $^{157}$Gd, have large cross sections of neutron captures to thermal neutrons 
	such as $^{155}$Gd(n,$\gamma$)$^{156}$Gd and $^{157}$Gd(n,$\gamma$)$^{158}$Gd.
	De-excitation gamma rays from $^{156}$Gd and $^{158}$Gd are self-absorbed in the scintillation crystals,
	and make spectral-line features which can be distinguished from other non-neutron signals. 
	The neutron burst lasted for $\sim$100~ms, and neutron fluences are estimated to be $>$52 and $>$31~neutrons~cm$^{-2}$ 
	at two observation points inside the power plant.
	Gadolinium orthosilicate scintillators work as valid detectors for thermal neutrons in lightning.

\end{abstract}

\maketitle


\section{Introduction}

	Since the first detection reported by Shah~et~al.\cite{Shah_1985}, 
	thunderstorms and lightning discharges have been thought to have an ability 
	to produce neutrons in the atmosphere\cite{Shyam_1999,Martin_2010,Gurevich_2012,Chilingarian_2012b,Gurevich_2015,Ishtiaq_2016,Kuroda_2016,Bowers_2017,Enoto_2017}.
	At first, neutrons were considered to be produced via deutron-deutron fusions $^{2}$H($^{2}$H,$n$)$^{3}$He 
	of vapor molecules in hot lightning paths\cite{Stephanakis_1972,Shah_1985,Ishtiaq_2016}.
	On the other hand, the discovery of high-energy phenomena in the atmosphere such as terrestrial gamma-ray flashes (TGFs) 
	have convinced that thunderstorms can produce neutrons 
	via photonuclear reactions with atmospheric nuclei\cite{Babich_2006,Babich_2007,Carlson_2010,Tavani_2011,Bowers_2017,Enoto_2017}.
	
	TGFs are brief bursts of energetic photons lasting for hundreds of microseconds, coincident with lightning discharges.
	They have been routinely detected by in-orbit gamma-ray monitors such as {\it Reuven Ramaty High Energy Solar Spectroscopic Imager}\cite{Smith_2005}, 
	{\it Astro‐Rivelatore Gamma a Immagini Leggero}\cite{Tavani_2011,Marisaldi_2010}, {\it Fermi}\cite{Briggs_2010,Mailyan_2016}, 
	and {\it Atmosphere-Space Interactions Monitor}\cite{Neubert_2019b,Ostgaard_2019b}, 
	after the discovery by {\it Compton Gamma-Ray Observatory}\cite{Fishman_1994}.
	Besides space-borne observations of upward-oriented TGFs, similar downward-oriented phenomena have been reported by ground-based experiments, which are now called 
	``downward TGFs''\cite{Dwyer_2004,Tran_2015,Hare_2016,Bowers_2017,Enoto_2017,Abbasi_2018,Smith_2018,Wada_2019b,Wada_2019c,Pleshinger_2019,Wada_2020}.
	Both upward and downward TGFs originate from bremsstrahlung of relativistic electrons accelerated and multiplied by high electric fields in lightning.	
	
	Energy spectra of TGFs were found to extend up to 40~MeV\cite{Smith_2005,Tavani_2011,Marisaldi_2010,Briggs_2011}.
	High-energy photons of $>$10~MeV can trigger photonuclear reactions with atmospheric nuclei 
	such as $^{14}$N($\gamma,n$)$^{13}$N (threshold 10.55~MeV) and $^{16}$O($\gamma,n$)$^{15}$O (15.66~MeV).
	Neutrons generated by photonuclear reactions have kinetic energies of MeV, 
	and are gradually thermalized in the atmosphere via multiple elastic scatterings with $^{14}$N\cite{Rutjes_2017,Bowers_2017,Enoto_2017}.
	When photoneutrons are generated at a low altitude, i.e. during low-charge-center winter thunderstorms, a part of neutrons arrives at the ground 
	while the rest is captured by ambient $^{14}$N via a neutron capture $^{14}$N($n,\gamma$)$^{15}$N or a charged-particle reaction $^{14}$N($n,p$)$^{14}$C\cite{Babich_2017}.
	Therefore, neutrons can be detected by ground-based apparatus in that case.

\begin{figure}[b!]
	\begin{center}
	\includegraphics[width=\hsize]{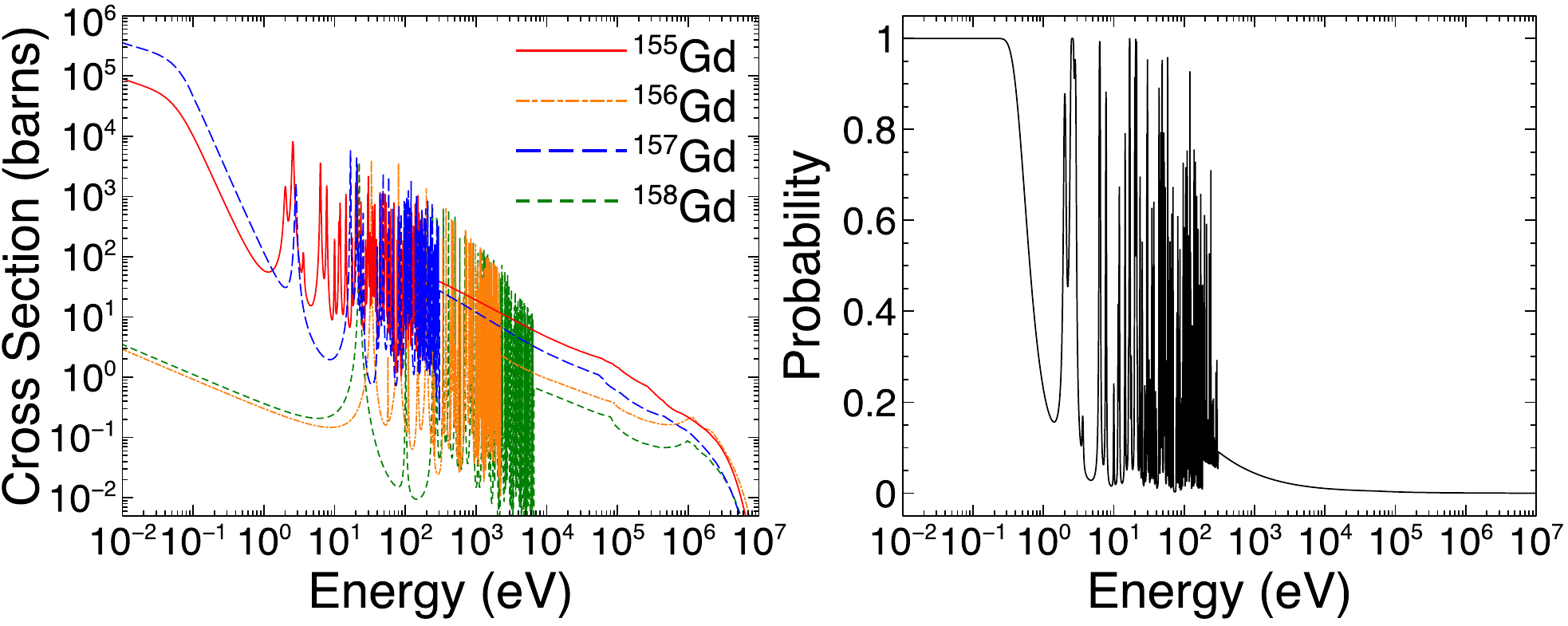}
	\caption{Figure~\ref{fig_cross_section}: Cross sections of neutron captures with gadolinium isotopes (left) 
	and a reaction probability of neutron captures with a 5-mm-thick GSO scintillator (right), 
	as functions of neutron kinetic energy. Line structures around keVs originate from resonance lines of Gd nuclei.
	Calculated with JENDL-4.0\cite{Shibata_2011}.}
	\label{fig_cross_section}
	\end{center}
\end{figure}

\begin{figure}[t!]
	\begin{center}
	\includegraphics[width=\hsize]{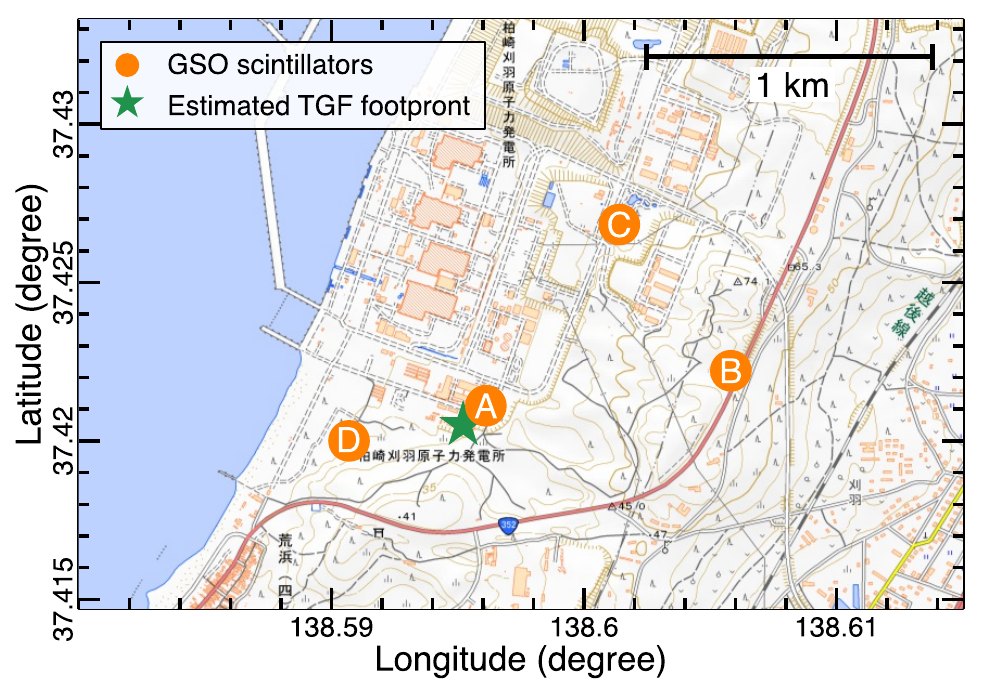}
	\caption{Figure~\ref{fig_map}: Locations of detectors (orange circles) at the Kashiwazaki-Kariwa nuclear power station.
	The TGF footprint (a green star marker) was estimated with an uncertainty of a 100-m radius, by our previous publication\cite{Wada_2019_prl}.}
	\label{fig_map}
	\end{center}
\end{figure}

	Detection techniques of thermal neutrons have been developed in various fields, such as astroparticle physics, nuclear security, non destructive inspection, etc.
	Common reactions utilized to detect thermal neutrons are neutron captures $^{1}$H($n,\gamma$)$^{2}$H, 
	$^{3}$He($n,p$)$^{2}$H, $^{6}$Li($n,\alpha$)$^{3}$He, and $^{10}$B($n,\alpha$)$^{7}$Li.
	For example, proportional counters filled with BF$_{3}$ (including $^{10}$B) or $^{3}$He gas detected thermal neutrons in lightning 
	at previous studies\cite{Shah_1985,Shyam_1999,Martin_2010,Gurevich_2012,Chilingarian_2012b,Ishtiaq_2016}.
	In addition to those reactions, gadolinium isotopes $^{155}$Gd and $^{157}$Gd have drawn attention 
	as another neutron detection scheme\cite{Kuroda_2012,Oguri_2014,Kuroda_2016,Marti-Magro_2017} thanks to their high cross sections to low-energy neutrons.
	Here we report a direct neutron detection by gadolinium orthosilicate scintillators coincident with a lightning discharge during a winter thunderstorm in Japan. 

\section{Instrument}

	Gadolinium orthosilicate scintillator (celium-doped Gd$_{2}$SiO$_{5}$: GSO) is a relatively new type of inorganic scintillation crystals.
	GSO is characterized by high density (6.7~g~cm$^{-3}$), fast decaying of scintillation light (40~ns), and high radiation resistance ($>10^{8}$~Gy).
	They were employed for the Hard X-ray Detector onboard the Japanese X-ray astronomy satellite {\it Suzaku}\cite{Takahashi_2007,Kokubun_2007}.
	
	Since GSO scintillators contain gadolinium isotopes, they are thought to be suitable for thermal neutron detection\cite{Reeder_1994}.
	The upper panel of Figure~\ref{fig_cross_section} shows cross sections of neutron captures with stable gadolinium isotopes.
	The isotopes $^{155}$Gd (14.8\% in nature) and $^{157}$Gd (15.7\%) have significantly high cross sections of neutron captures 
	to thermal neutrons (0.025~eV) of $6.1 \times 10^{4}$ and $2.5 \times 10^{5}$~barns, respectively.
	As shown in the lower panel of Figure~\ref{fig_cross_section}, a 5-mm thick GSO scintillator stops almost all neutrons below 0.3~eV
	via neutron captures $^{155}$Gd($n$,$\gamma$)$^{156}$Gd and $^{157}$Gd($n$,$\gamma$)$^{158}$Gd.
	After Gd nuclei capture a neutron, they emit de-excitation gamma rays; 
	$^{156}$Gd mainly emits gamma-ray lines at 88.97 and 199.22~keV, and $^{158}$Gd at 79.51 and 181.94~keV\cite{Shibata_2011}. 
	The de-excitation lines are self-absorbed in the GSO scintillators, 
	and hence they make a clear spectral-line feature for neutron detection.	

	The Gamma-ray Observation of Winter Thunderclouds (GROWTH) experiment has been successfully operated in coastal areas of the Sea of Japan 
	since 2006\cite{Tsuchiya_2007,Tsuchiya_2011,Tsuchiya_2013,Umemoto_2016,Enoto_2017,Wada_2018,Wada_2019_commphys,Wada_2019_prl}.
	One of our observation sites, the Kashiwazaki-Kariwa nuclear power station of Tokyo Electric Power Company Holdings in Niigata Prefecture, Japan, 
	was upgraded with four gamma-ray detectors in 2016.
	The locations of the gamma-ray detectors are shown in Figure~\ref{fig_map}.
	Based on the discovery of photoneutron productions in winter lightning\cite{Enoto_2017,Bowers_2017}, 
	GSO scintillators were added to the four detectors in 2017 for neutron detection.
	We utilized GSO scintillators of $2.4 \times 2.4 \times 0.5$~cm$^{3}$.
	These are connected with a photo-multiplier tube of Hamamatsu R7600U, and read out 
	by our original data acquisition system\cite{Enoto_2017,Wada_2018,Wada_2019_tepa}.

\section{Calibration in laboratory}

\begin{figure}[t]
	\begin{center}
	\includegraphics[width=\hsize]{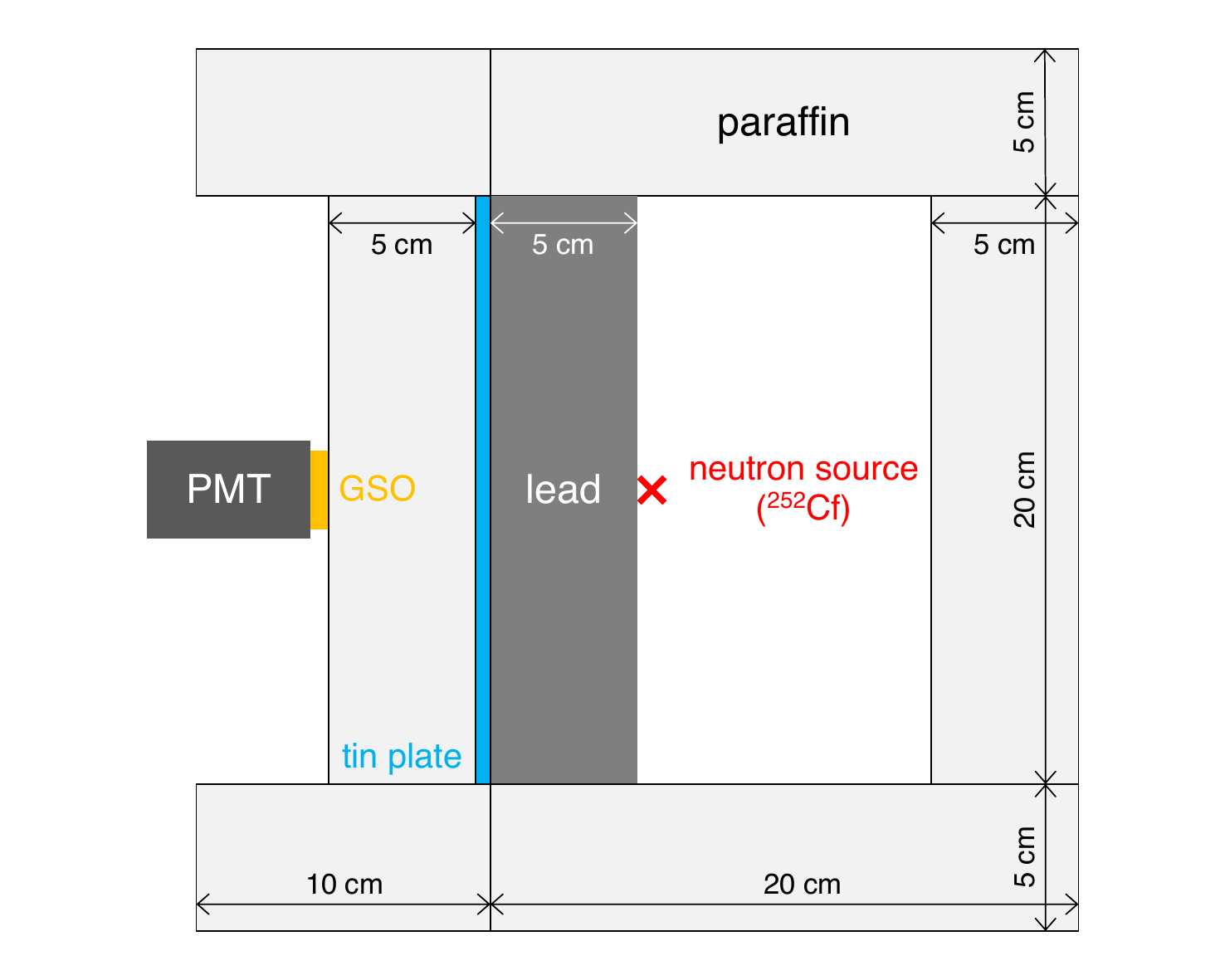}
	\caption{Figure~\ref{fig_calibration_setup}: A schematic view of the experiment setup for the GSO calibration.}
	\label{fig_calibration_setup}
	\end{center}
\end{figure}

\begin{figure}[t]
	\begin{center}
	\includegraphics[width=\hsize]{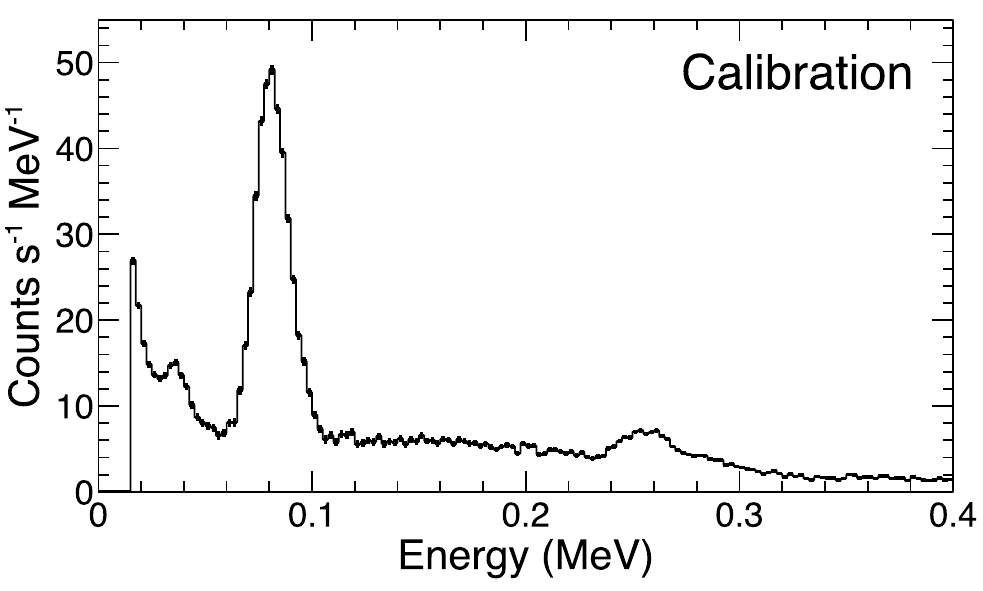}
	\caption{Figure~\ref{fig_calibration_spec}: A background-subtracted spectrum of neutron captures in the GSO scintillator measured with the calibration setup.}
	\label{fig_calibration_spec}
	\end{center}
\end{figure}

    This laboratory calibration aims at confirming spectral features of neutron captures by Gd nuclei, 
    and constraining a conversion factor to estimate the number of neutron captures in the GSO scintillators 
    from intensities of de-excitation lines.
	The intensity of de-excitation lines is affected by various processes: branching ratios of de-excitation lines, 
	detection efficiency of gamma-ray photons inside scintillators, 
	and simultaneous self-absorption of multiple de-excitation lines. To take these effects into account, 
	we performed a calibration measurement by irradiating neutrons to a GSO scintillator.
	We utilized $^{252}$Cf as a neutron source, which exhibits spontaneous fissions with a half life of 2.645~years; 0.188~neutrons are emitted per a decay on average. 
	The energy spectrum of neutrons emitted from this isotope follows $E^{0.5} \exp(-E / 1.656~{\rm MeV})$, where $E$ is the kinetic energy of neutrons\citep{Meadows_1967}.
	The $^{252}$Cf source utilized here had a radioactivity of 30~kBq at the moment of the measurement, calibrated by the manufacturer of this source; 
	$3.5 \times 10^{3}$~neutrons were emitted per second. 
	Note that 30\% systematic uncertainty is claimed to the radioactivity.

	The measurement setup is shown in Figure~\ref{fig_calibration_setup}.
	A lead block of 5-cm thickness, a tin plate of 3-mm thickness, and a paraffin block of 5-cm thickness are placed between the GSO scintillator and the neutron source.
	The lead block reduces background counts in GSO by screening gamma rays from $^{252}$Cf.
	Neutrons penetrating the lead block are thermalized by the paraffin block, then enter the GSO as thermal or epithermal neutrons.
	When the lead block absorbs gamma rays, the K$\alpha$ X-ray line at 74.2~keV can be emitted.
	This line contaminates the energy spectrum in GSO and be mixed up with 89.0~keV and 79.5~keV gamma rays from $^{156}$Gd and $^{158}$Gd respectively, because the energy resolution of this setup at is 14.3~keV at 81~keV (17.6\%; FWHM).
	Therefore, the tin plate is inserted to screen the K$\alpha$ line from lead. 
	The plate of 3-mm thickness cuts 99.8\% of 80-keV photons. X-rays from tin (at 25.2~keV) do not matter in this measurement.
	Energy calibration of GSO was performed with the 32- and 662-keV lines of $^{137}$Cs, 81- and 356-keV lines of $^{133}$Ba.
	The gain calibration accuracy is within 4\%, which is considered as a systematic error.
	
	Measurements with $^{252}$Cf and background measurements were performed for 45.5 hours and 124 hours, respectively. 
	Figure~\ref{fig_calibration_spec} presents the obtained energy spectrum. The most significant line is at $\sim$80~keV. 
	In addition, lines around 35~keV and 260~keV are also found. These spectral features are consistent with a previous work\citep{Reeder_1994}.
	By evaluating the primary line with a Gaussian and a continuum component, the center and count rate of the line is determined 
	to be $81.08 \pm 0.08~({\rm stat.}) \pm 3.20~({\rm sys.})$~keV and $0.794 \pm 0.009~{\rm count}~{\rm s}^{-1}$, respectively.
	Statistical uncertainties shown in the present paper are at 1$\sigma$ confidence level. 
	The line center is consistent with the 79.5~keV line from $^{158}$Gd. Therefore, the line mainly originates from neutron captures $^{157}$Gd($n$,$\gamma$)$^{158}$Gd.
	It is thought that the contribution from $^{155}$Gd($n$,$\gamma$)$^{156}$Gd, which emits a 89.0~keV line, is smaller than $^{157}$Gd($n$,$\gamma$)$^{158}$Gd 
	because its cross section to thermal neutrons is one forth of $^{157}$Gd($n$,$\gamma$)$^{158}$Gd.
	In the same way, the center of the $\sim$260-keV line is determined to be $258.6 \pm 1.1 ({\rm stat.}) \pm 10.4 ({\rm sys.})$~keV.
	This line is consistent with a simultaneous detection of 79.5-keV and 181.9-keV lines from $^{158}$Gd as one line at 261.4~keV.
	In addition, $^{156}$Gd and $^{158}$Gd emit 38.7-keV and 29.3-keV electrons by internal conversions instead of 89.0-keV and 79.5-keV gamma rays, respectively\citep{Reeder_1994}.
	The line structure around 35~keV seems to originate from monochromatic electrons of the internal conversion.
	
	A Monte-Carlo simulation was then performed to test the number of neutrons captured in GSO in the geometry of the present experiment.
	A mass model of the geometry shown in Figure~\ref{fig_calibration_setup} is implemented in the simulation.
	Neutrons with the spectrum from $^{252}$Cf fissions were generated isotropically,
	then the number of the reactions $^{155}$Gd(n,$\gamma$)$^{156}$Gd and $^{157}$Gd(n,$\gamma$)$^{158}$Gd is registered.
	When neutrons are captured in GSO, tracking of their secondary products was terminated.
	Here we employed the neutron cross-section database JENDL-4.0\citep{Shibata_2011}, developed and distributed by Japan Atomic Energy Agency.
	
	When $10^{9}$ neutrons were generated in the simulation, $2.16 \times 10^{5}$ reactions of $^{155}$Gd(n,$\gamma$)$^{156}$Gd 
	and $7.45 \times 10^{5}$ reactions of $^{157}$Gd(n,$\gamma$)$^{158}$Gd were registered. 
	For the present geometry, the ratios of the reactions $^{155}$Gd(n,$\gamma$)$^{156}$Gd and $^{157}$Gd(n,$\gamma$)$^{158}$Gd 
	to the total number of the generated neutrons are 0.022\% and 0.075\% respectively, and 0.097\% in total.

	Then, the simulation and the measurement are compared. 
	The neutron source $^{252}$Cf emitted $(3.5 \pm 1.1) \times 10^{3}~{\rm neutrons}~{\rm s}^{-1}$ at the moment of the calibration measurement.
	Combining the neutron-emission rate with the ratio 0.097\% obtained by the simulation, 
	an expected neutron-capture rate in the GSO scintillator is $3.4 \pm 1.1~{\rm neutrons}~{\rm s}^{-1}$.
	For comparison, the calibration measurement derived that the main 80-keV peak in Figure~\ref{fig_calibration_spec} has an intensity of $0.794 \pm 0.009~{\rm count}~{\rm s}^{-1}$.
	Therefore, one neutron-capture reaction inside a GSO scintillator makes $0.23 \pm 0.08$ counts at 80-keV.
	We adopted this number as a conversion factor to estimate neutron fluences in the following sections. 

\begin{figure}[h]
	\begin{center}
	\includegraphics[width=\hsize]{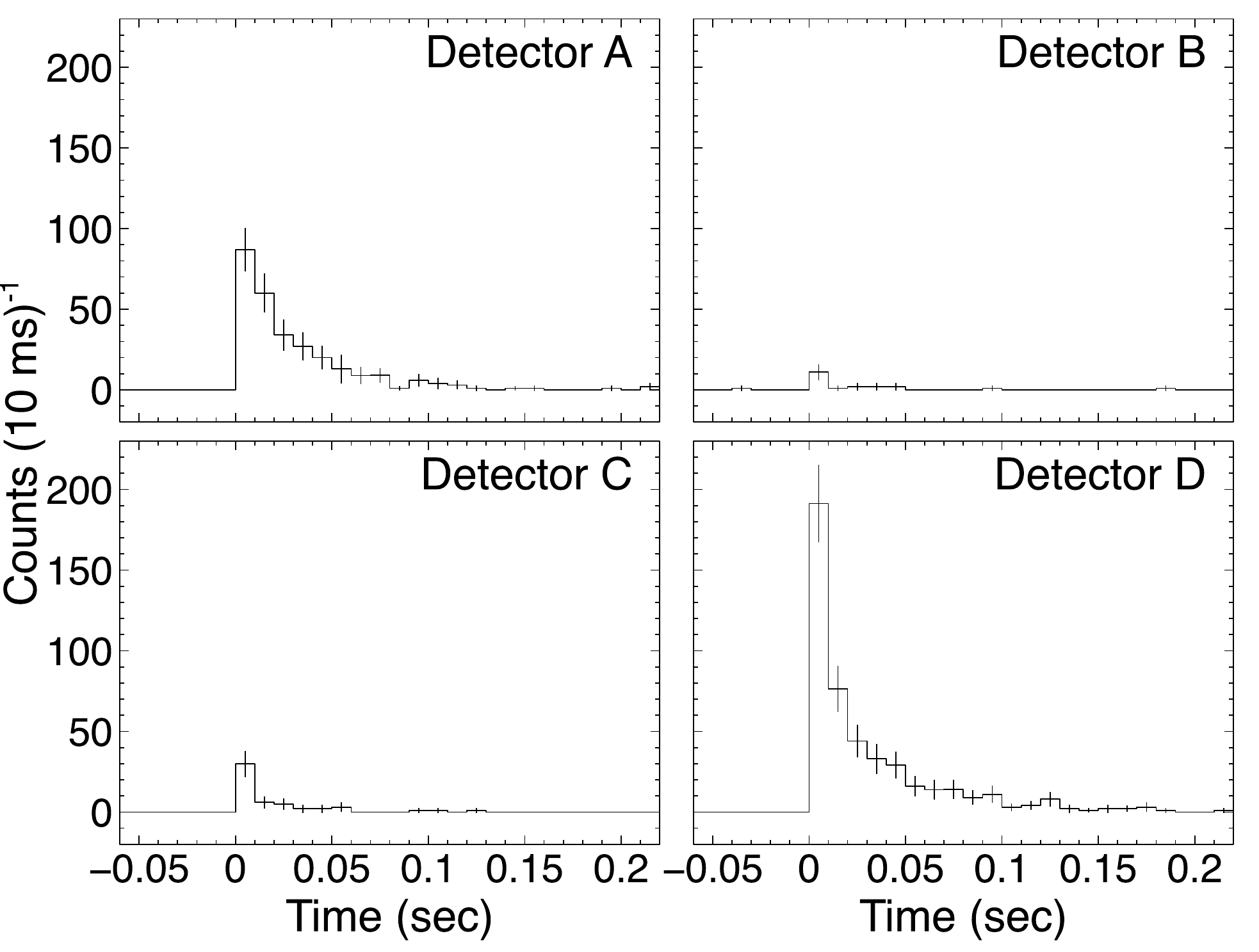}
	\caption{Figure~\ref{fig_lightcurve}: Count-rate histories in 0.04--1.0~MeV obtained with GSO scintillators. The origin of time is the beginning of the downward TGF.}
	\label{fig_lightcurve}
	\end{center}
\end{figure}

\begin{figure}[tbh]
	\begin{center}
	\includegraphics[width=\hsize]{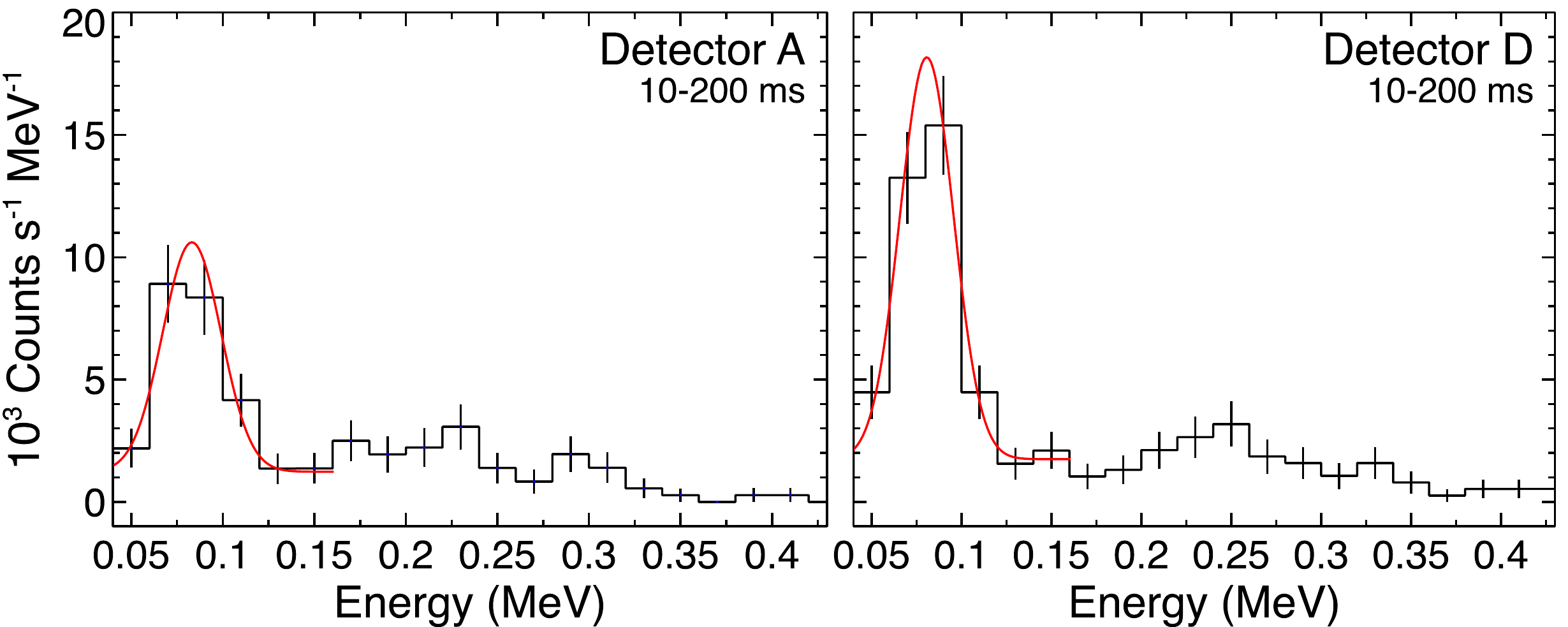}
	\caption{Figure~\ref{fig_spectrum}: Background-subtracted spectra of GSO scintillators. 
	The overlaid red lines present the best-fit models of a line structure around 80~keV.
	Background spectra were accumulated for 10~minutes before the downward TGF.}
	\label{fig_spectrum}
	\end{center}
\end{figure}

\section{Observation}

	At 10:03:02, November 24th, 2017 (in Coordinated Universal Time), our gamma-ray detectors and monitoring posts operated by the power station 
	recorded a downward TGF, as we previously reported\cite{Wada_2019_prl}.
    The downward TGF was followed by de-excitation gamma rays of neutron captures in the atmosphere, originating from photonuclear reactions.
	At the same time as the detection of the downward TGF and the photonuclear reactions, 
	the GSO scintillators also recorded an increase in count rates lasting for $\sim$100~ms.
	Count-rate histories obtained by the GSO scintillators are shown in Figure~\ref{fig_lightcurve}.
	Significant increases in count rates were observed by Detectors~A and~D coincident with the lightning discharge.

	Energy spectra recorded by Detectors~A and~D were extracted from 10~ms to 200~ms after the lightning discharge, and presented in Figure~\ref{fig_spectrum}.
	Since the initial 10~ms was disturbed by the downward TGF itself, this time domain was excluded for spectral analysis.
	Both spectra have a significant line feature at a low energy range around 80~keV. 
	The center energy of the line was evaluated as $83.1 \pm 2.8~({\rm stat.}) \pm 3.2~({\rm sys.})$~keV 
	and $80.7 \pm 1.9~({\rm stat.}) \pm 3.2~({\rm sys.})$~keV for Detectors~A and~D respectively by fitting with a Gaussian function plus a constant component.
	This is consistent with the center energy at 81~keV, obtained by the calibration measurement.
	Therefore, this is a successful detection of neutrons by GSO scintillators. 
	
	The photon counts at the line were also evaluated to be $71 \pm 18$~counts and $116 \pm 23$~counts for Detectors~A and~D respectively by the spectral fitting.
	By utilizing the conversion factor $0.23 \pm 0.08$~counts per one neutron capture obtained by the calibration,
	$(3.1 \pm 1.3) \times 10^{2}$ and $(5.0 \pm 2.0) \times 10^{2}$ neutrons were captured in the GSO scintillators of Detectors~A and D, respectively.

\section{Discussion}

	The GSO scintillators employed in the present study have a detection area of $2.4 \times 2.4$~cm$^{2}$.
	For thermal neutrons, whose kinetic energies are 0.025~eV or less, the detection efficiency is almost 1.0 (Figure~\ref{fig_cross_section}),
	and the effective area of the GSO scintillators is 5.76~cm$^{2}$.
	In the actual situation, however, neutrons are not totally thermalized, 
	and epi-thermal and fast neutrons must be also reaching the ground (e.g. Figure~3 in Bowers et al.\cite{Bowers_2017}), 
	not interacting with GSO (e.g. Figure~3 in Bowers et al.\cite{Bowers_2017})
	Therefore, we can only estimate lower limits of neutron fluences on the ground, based on the recorded number of neutron captures;
	$>$31~neutrons~cm$^{-2}$ and $>$52~neutrons~cm$^{-2}$ for Detectors~A and~D, respectively.

	In our previous publicaiton\cite{Wada_2019_prl}, we estimated the height, position, and the number of avalanche electrons of the downward TGF 
	based on the on-ground measurement of radiation doses by monitoring posts.
	The footprint of the downward TGF was located 100~m southwest from Detector~A, as shown in Figure~\ref{fig_map}.
	In the present result, GSO scintillators of Detectors~A and~D observed a significant number of neutrons, while Detectors~B and~C did not.
	Therefore, a larger number of neutrons were generated by the downward TGF around Detectors~A and~D, rather than around Detectors~B and~C.
	This is consistent with our previous estimation of the footprint\cite{Wada_2019_prl}.
	The height and the number of avalanche electrons of the downward TGF 
	had been also estimated to be $2.5 \pm 0.5$~km and $8^{+8}_{-4} \times 10^{18}$~electrons (above 1~MeV)\cite{Wada_2019_prl}.
	To compare the estimation and the present result of neutron fluences, we need end-to-end Monte-Carlo simulations
	calculating photonuclear reactions and propagation of neutrons in the atmosphere, which will be covered as a future work.

	This paper presents that neutrons reaching the ground were directly detected by GSO scintillators coincident with a lightning discharge.
	Besides neutrons, de-excitation gamma-ray photons via atmospheric neutron captures $^{14}$N($n,\gamma$)$^{15}$N also reached the ground simultaneously.
	In even such a high-radiation environment, GSO scintillators work as valid detectors for thermal neutron, 
	as de-excitation gamma-ray lines of neutron captures with gadolinium isotopes were self-absorbed and clearly identified in energy spectra.

\begin{acknowledgments}

	We deeply thank the radiation safety group of the Kashiwazaki-Kariwa nuclear power plant for providing the observation site.
	Monte-Carlo simulations were performed on the Hokusai BigWaterfall supercomputing system 
	operated by RIKEN Head Office for Information Systems and Cybersecurity.  
	This research is supported by JSPS/MEXT KAKENHI grants 16H06006, 17K05659, 18J13355, 18H01236, 19H00683,
	and citizen supporters via an academic crowdfunding platform ``academist''.
	T.E. is supported by Kyoto University and RIKEN Hakubi Fellow Programs.
	The background image in Figure~\ref{fig_map} was provided by the Geospatial Information Authority of Japan.

\end{acknowledgments}



%



\end{document}